\documentclass[12pt]{article}

\usepackage{amssymb}
\usepackage{amsmath}
\usepackage{graphicx}

\catcode`@=11
\renewcommand{\section}{\@startsection{section}{1}{0pt}{\medskipamount}
{\medskipamount}{\large\bf}}
\numberwithin{equation}{section}
\catcode`@=12

\def\beq{\begin{eqnarray}}    
\def\eeq{\end{eqnarray}}      

\def\ln{\,\mbox{ln}\,}                  

\def\diff{\textrm{d}}
\def\sfrac#1#2{{\textstyle\frac{#1}{#2}}}
\def\={\ =\ }

\begin{document}

\begin{titlepage}
\setcounter{page}{0}


\begin{center}

{\LARGE\bf
Remarks on the Curci-Ferrari  model}

\vspace{18mm}

{\Large
Peter M. Lavrov 
$\footnote{E-mail: lavrov@tspu.edu.ru}$,
}

\vspace{8mm}

\noindent 
{\em Tomsk State Pedagogical University,\\
Kievskaya St.\ 60, 634061 Tomsk, Russia}

\end{center}
\vspace{18mm}

\begin{abstract}
\noindent Dependence of Green's functions for the Curci-Ferrari
model  on the parameter resembling the gauge parameter in massless
Yang-Mills theories is investigated. It is shown that the generating
functional of vertex functions  (effective action)   depends on
this parameter on-shell.
\end{abstract}

\vfill


\end{titlepage}


\section{Introduction}

\noindent Recently,  it was claimed \cite{Kondo} that the
Curci-Ferrari (CF) model \cite{CF1} can be presented as a unitary
and renormalizable model for massive Yang-Mills fields without Higgs
fields. From the beginning it was well known that the CF model obeys
the property of renormalizability \cite{CF2,dBSvNW} and the action
of this model is invariant under modified BRST and anti-BRST
transformations. In massless limit, the action of the CF model
reduces to the Faddeev-Popov (FP) action \cite{FP} constructed in a
one parameter linear gauge. The FP action is invariant under the
BRST transformations \cite{brs,t} as well as under the anti-BRST
transformations in special gauges \cite{CF3,Oj}. The BRST symmetry
plays a fundamental role in quantum theory of gauge fields
\cite{NOj}.  Note, for example, that breaking of BRST symmetry as it
occurs in Yang-Mills theories when one takes into account the Gribov
horizon \cite{Gribov,Zwanziger1,Zwanziger2} leads to the gauge
dependence of effective action in gauge theories on-shell
\cite{llr,lrr}. In turn it means inconsistency for physical
interpretation of results obtained within this approach. In
Yang-Mills theories both the BRST and anti-BRST transformations are
nilpotent. Nilpotency of the BRST symmetry allows to formulate
suitable conditions (the so-called Kugo-Ojima criterion) for a
physical state space providing unitarity of S-matrix in non-abelian
gauge theories \cite{KO}. In contrast to this case, the modified
BRST and anti-BRST transformations in the CF model are not
nilpotent. Namely, this fact was considered as a reason for
violation of unitarity in this theory for a long time
\cite{CF2,Oj1,dBSvNW}. Reformulation of the CF model proposed in
\cite{Kondo} is connected with using  local non-linear
transformations of massive vector fields and rewritten the CF action
in terms of new variables to obtain a model for massive Yang-Mills
fields without Higgs fields. The statement about unitarity of
S-matrix for this model contradicts with previous conclusions about
non-unitarity of the CF model \cite{CF2,Oj1,dBSvNW} and sounds
rather strange from the point of view of the equivalence theorem
\cite{KT,T} because two theories under consideration are connected
through a change of variables which satisfies conditions of the
theorem. Now it is clear that the unitarity problem for the CF model
\cite{CF1} and the model of massive Yang-Mills fields without Higgs
fields \cite{Kondo} needs in further investigations.

In present paper  the dependence of Green's functions for the CF
model on a parameter  resembling the gauge
parameter in massless Yang-Mills theories is investigated. It is
shown that the generating functional of vertex functions (effective
action) does  depend on this parameter even on-shell.

The paper is organized as follows. In Section~2, the CF model is
considered. In Section 3, dependence of Green's functions  for the
CF model on the parameter $\beta$ is studied. Finally, Section 4
gives concluding remarks.

We employ the condensed notation of DeWitt~\cite{DeWitt}.
Derivatives with respect to sources  are taken from the left,
while those with respect to fields are taken from the right.
Left derivatives with respect to fields are labeled by a subscript~$l$.

\section{The Curci-Ferrari model}
\noindent Consider a massive extension of the massless Yang-Mills
theory  proposed by Curci and Ferrari \cite{CF1}. The CF
model is described by the action
\beq
\label{CFa}
S=S_{YM}+S_{gf}+S_m,
\eeq
where $S_{YM}$ is the Yang-Mills action of fields $A_{\mu}^a$, which
take values in the adjoint representation of~the Lie algebra $su(N)$ so that,
$a=1,\ldots,N^2{-}1$,
\beq
S_{YM}=-\frac{1}{4} F_{\mu\nu}^{a}F^{a\mu\nu}
\qquad\textrm{with}\quad
F^a_{\mu\nu}=\partial_{\mu}A^a_{\nu}-\partial_{\nu}A^a_{\mu}+
f^{abc}A^b_{\mu}A^c_{\nu}\ , \label{clYM}
\eeq
and $\mu,\nu=0,1,\ldots,D{-}1$, the Minkowski space has signature
$(+,-,\ldots,-)$, and $f^{abc}$ denote the (totally antisymmetric) structure
constants of $su(N)$, the symbol $\int\!\diff^D x$ is omitted.
The action $S_{gf}$ has the form
\beq \label{agf} S_{gf}=
B^a\partial^{\mu}A^a_{\mu} +{\bar
C}^a\partial^{\mu}D^{ab}_{\mu}C^b+\frac{\beta}{4}B^aB^a+\frac{\beta}{4}{\bar
B}^a{\bar B}^a, \eeq
where
\beq \label{N} {\bar B}^a=-B^a+N^a, \quad N^a=N^a(C,{\bar C})=
f^{abc}{\bar C}^bC^a,\quad D^{ab}_{\mu}=
\delta^{ab}\partial_{\mu}+f^{acb}A^c_{\mu} \eeq
and $\beta$ is a
parameter of the model. The action $S_m$ contains a mass $m$ for the
vector fields $A^a_{\mu}$ and the ghosts $C^a$ and antighosts ${\bar
C}^a$
\beq \label{am}
S_m=\frac{1}{2}m^2
A^a_{\mu}A^{a\mu}+\beta m^2 {\bar C}^aC^a.
\eeq
Here the notations $B^a$ for bosonic auxiliary fields were used.  In
massless limit they are identified with the Nakanishi - Lautrup
fields.

Note that $S_{YM}+S_{gf}$ can be presented as the action constructed
by the rules of Faddeev-Popov quantization \cite{FP} , $S_{FP}$, in
one-parameter linear gauge $\chi^a$
\beq \label{gauge}
\chi^a=\partial^{\mu}A^a_{\mu}+\frac{\beta}{2}B^a \eeq
and modified by the additional term $S_{ad}$
\beq
S_{YM}+S_{gf}=S_{FP}+S_{ad},
\eeq
where
\beq S_{ad}=\frac{\beta}{4}N^aN^a-
\frac{\beta}{2}B^aN^a.
\eeq

The action (\ref{clYM}) is invariant under the gauge transformations
\beq \delta A^a_{\mu}=D^{ab}_{\mu}\xi^b ,\eeq
where $\xi^a=\xi^a(x)$ are arbitrary functions of space-time coordinates.
In turn, the actions $S_{FP}$ and $S_{ad}$ are invariant  under
BRST transformation \cite{brs,t}
\beq\label{BRSTtr}
\nonumber
\delta_B A_{\mu}^{a} &=& D^{ab}_{\mu}C^b\theta ,\\
\delta_B C^a &=& \frac{1}{2} f^{abc}C^bC^c\theta ,\\
\nonumber
\delta_B \bar{C}{}^a &=& B^a\theta ,\\
\nonumber
\delta_B B^a &=& 0\ ,
\label{BRSTGZred} \eeq
where $\theta$ is a constant Grassmann parameter. Moreover,  these
actions are invariant under the anti-BRST transformation
\cite{CF3,Oj}
\beq\label{antiBRSTtr}
\nonumber
{\bar \delta}_B A_{\mu}^{a} &=&
D^{ab}_{\mu}{\bar C}^b{\bar \theta} ,\\
{\bar \delta}_B C^a &=& (-B^a+ f^{abc}{\bar C}^bC^c){\bar \theta},\\
\nonumber
 {\bar \delta}_B {\bar
C}^a &=& \frac{1}{2} f^{abc}{\bar C}^b{\bar C}^c{\bar \theta} , \\
\nonumber
{\bar \delta}_B B^a &=&-f^{abc}{\bar C}^bB^c {\bar \theta}\ ,
 \eeq
with  ${\bar\theta}$ being a constant Grassmann parameter
\cite{CF1}. The action of the CF model is not invariant under the
BRST transformation because of $\delta_B S_m\neq 0$ but it is
invariant under the modified BRST transformation $\delta_{mB}S=0$
\cite{CF1}, where
\beq
\label{mBRSTtr}
\nonumber
\delta_{mB} A_{\mu}^{a} &=& D^{ab}_{\mu}C^b\theta\ ,\\
\delta_{mB} C^a &=& \frac{1}{2} f^{abc}C^bC^c\theta,\\
\nonumber
\delta_{mB} \bar{C}{}^a &=& B^a\theta\ ,\\
\nonumber
\delta_{mB} B^a &=& m^2C^a\theta,
\eeq
as well as under the modified anti-BRST transformation  ${\bar
\delta}_{mB}S=0$ \cite{CF1}. In what follows the explicit form of
the modified anti-BRST transformation will not be essential, and we
omit it. Note only that existence of anti-BRST symmetry for
Yang-Mills theories in the gauge (\ref{gauge}) is not specific property of
these theories in special gauges. For any classical gauge theory in
any admissible gauge one can construct a quantum version respecting
both the BRST and anti-BRST symmetries \cite{blt1,blt2,blt3,l}. In
contrast to the usual BRST (or anti-BRST) transformation, the
modified BRST (or modified anti-BRST) transformation is not
nilpotent. It was a reason to claim violation of unitarity for the
CF model \cite{CF2,Oj1,dBSvNW}.

Returning to the CF model it needs definitely  to say that from the
beginning it should be considered as a non-gauge model in contrast
to the Faddeev-Popov action $S_{FP}$ constructed for Yang-Mills
action $S_{YM}$ which is invariant under gauge transformations
$\delta A^a_{\mu}=D^{ab}_{\mu}\xi^b$. In sector of vector fields
$A^a_{\mu}$ the action of CF model, $S_{mYM}$,
\beq\nonumber S_{mYM}=
-\frac{1}{4}F_{\mu\nu}^{a}F^{\mu\nu{}a}+\frac{1}{2}
m^2A^a_{\mu}A^{a\mu} \eeq
is not gauge invariant at all. In particular, there is no reason to
refer $\beta$ as the gauge parameter. It is a parameter of the
theory with initial classical non-degenerated action $S$
(\ref{CFa}), for which the physical space contains particles
corresponding to the massive vector fields $A^a_{\mu}$ and scalar
anticommuting fields $C^a, {\bar C}^a$. This point of view will be supported
in the next Section by investigation of the dependence of Green's functions on
this parameter.

\section{Dependence of Green's functions on parameter $\beta$}

\noindent In this section we will study dependence of Green's
functions for the CF model (\ref{CFa})-(\ref{am}) on parameter
$\beta$. We start with the vacuum functional $Z_{\beta}$ for the CF
model explicitly indicating dependence on $\beta$
\beq Z_{\beta}=\int D\phi \exp\Big(\frac{i}{\hbar}\;S\;\Big), \eeq
where $\phi$ denotes the set of all fields of the theory under
consideration,
\beq
\label{phi}
\phi^i=(A^a_{\mu}, {\bar C}^a, C^a,B^a).
\eeq
Let $Z_{\beta+\delta\beta}$ be the vacuum functional corresponding to
small variation of the parameter $\beta$:
$\beta\rightarrow\beta+\delta\beta$. It leads to variation of the CF action
(\ref{CFa}): $S\rightarrow S+\delta_{\beta}S$,
where
\beq
\label{dS}
\delta_{\beta}S=\Big(\frac{1}{4}B^aB^a+
\frac{1}{4}{\bar B}^a{\bar B^a}+m^2{\bar C}^aC^a\Big)\delta\beta.
\eeq
Then we have
\beq
\label{vZ}
Z_{\beta+\delta\beta}=\int D\phi \exp\Big(\frac{i}{\hbar}\Big[\;S\;+
\delta_{\beta}S\Big]\Big).
\eeq
From (\ref{dS}) and (\ref{vZ}) it follows the equation
\beq
\label{Zind1}
\frac{\partial Z_{\beta}}{\partial\beta}=\frac{i}{4\hbar}<B^aB^a> +
\frac{i}{4\hbar}<{\bar B}^a{\bar B^a}>+\frac{i}{\hbar}m^2<{\bar C}^aC^a>
\eeq
where $<\cdots>$ means a vacuum expectation  value of corresponding
quantities, for example,
\beq
<{\bar C}^aC^a>=\int D\phi \;{\bar C}^aC^a\;\exp\Big(\frac{i}{\hbar}\;S\;\Big).
\eeq
Now let us use the invariance of $S$ (\ref{CFa})  under the modified
BRST transformation (\ref{mBRSTtr}) to investigate the functional
$Z_{\beta+\delta\beta}$.  To this end, in the functional integral
(\ref{vZ}) we can make a change of variables being given by Eqs.
(\ref{mBRSTtr}) with some functional $\Lambda=\Lambda(\phi)$ instead
of the constant Grassmann odd parameter $\theta$. It is clear that
the CF action (\ref{CFa}) is invariant under such a change of
variables. If we restrict ourself to the first order in
$\Lambda(\phi)$ and $\delta\beta$ then there appears contribution
only coming from the integration measure
\beq
\label{vZch}
Z_{\beta+\delta\beta}=\int D\phi \exp\Big(\frac{i}{\hbar}\Big[\;S\;+
\delta_{\beta}S+\delta M\Big]\Big),
\eeq
where
\beq
\label{Mch}
\delta M=-i\hbar  \Big(\frac{\delta
\Lambda(\phi)}{\delta
A^a_{\mu}}D^{ab}_{\mu}C^b-\frac{1}{2}\frac{\delta\Lambda(\phi)}{\delta
C^a}f^{abc}C^bC^c- \frac{\delta\Lambda(\phi)}{\delta {\bar C}^a}B^a-
m^2\frac{\delta\Lambda(\phi)}{\delta B^a}C^a\Big). \eeq
Choosing the functional $\Lambda(\phi)$ as
\beq
\label{Lf}
\Lambda(\phi)=\frac{i}{2\hbar}\Big({\bar C}^aB^a-
\frac{1}{2} f^{abc}{\bar C}^a{\bar C}^bC^c\Big)\delta\beta
\eeq
we find that
\beq
\label{nonpertur}
\delta_{\beta}S+\delta M=\frac{1}{2}m^2{\bar C}^aC^a\delta\beta.
\eeq

In massless limit, the vacuum functional $Z_{\beta}$ does not depend
on the parameter $\beta$. It is no wonder that there is no
dependence on this parameter because in this limit the CF action
reduces to the FP action for massless Yang-Mills when $\beta$ plays
a role of gauge parameter and nilpotency of the BRST transformations
is restored. If $m\neq 0$ then there is an essential dependence of
vacuum functional on this parameter and $\beta$ becomes a physical
parameter defining, for example, a mass, $m_c$, of scalar
anticommuting fields $C^a$ and ${\bar C}^a$ in the form $m^2_c=\beta
m^2$ because the equations of motion read
\beq
\nonumber
(\square +m^2_c)C^a+\cdots =0,
\eeq
where $\square =\partial^{\mu}\partial_{\mu}$ and the dots mean
terms which are non-linear in $\phi^i$. Similar equations hold for
fields ${\bar C}^a$. Unfortunately, we cannot use the relation
(\ref{nonpertur}) to find the representation of dependence of
$Z_{\beta}$ on $\beta$
\beq
\nonumber
\label{Zind}
\frac{\partial Z_{\beta}}{\partial\beta}=\frac{i}{2\hbar}m^2<{\bar C}^aC^a>
\eeq
as one might think considering (\ref{vZch}) and (\ref{nonpertur}).
In the case $m^2\neq 0$ the dependence of $Z_{\beta}$ on $\beta$
becomes essential and the change of variables
\beq \nonumber \phi^i\rightarrow \phi^{'i}=\phi^i+
\frac{i}{\hbar}{\bar \Lambda}(\phi)R^i(\phi),\quad
\Lambda(\phi)=\frac{i}{\hbar}{\bar \Lambda}(\phi) \eeq
used in (\ref{Mch}) and (\ref{Lf}) is beyond the strong definition
of functional integral within loop expansions (in $\hbar$)
\cite{Slavnov}. Here the condensed notations
$\delta_{mB}\phi^i=R^i(\phi)\theta$ for the modified BRST
transformation (\ref{mBRSTtr}) were used. Note that such kind of
transformations serves as a tool to prove the gauge independence of
vacuum functional (and physical quantities) in Yang-Mills theories
as well as in general gauge theories \cite{BV}. In the case of gauge
theories, it does not lead to conflicts if one considers physical
quantities because they are gauge invariant ones and the change of
variables touches  a modification of gauge fixing functional only.

We can investigate dependence of Green's functions on parameter
$\beta$ for the CF model as well. The generating functional of
Green's functions, $Z_{\beta}(J)$, is written in the form
\beq
Z_{\beta}(J)=\int D\phi \exp\Big(\frac{i}{\hbar}[S(\phi)+J_i\phi^i]\Big),
\eeq
where the action $S$  is defined  through relations
(\ref{CFa})-(\ref{am}), the set of fields $\phi^i$ is given in
(\ref{phi}) and $J_i=(j^{a}_{\mu}, K^a,{\bar K}^a, L^a)$ are usual
sources to fields $\phi^i$ with relevant distributions of Grassmann
and ghost parities. Let us consider the CF model corresponding a
small variation of parameter $\beta$ ( $\beta\rightarrow
\beta+\delta\beta$). Then the generating functional for Green's
functions is
\beq \label{Zbd} Z_{\beta+\delta\beta}(J)=\int D\phi
\exp\Big(\frac{i}{\hbar}[S(\phi)+\delta_{\beta}S+J_i\phi^i]\Big)
\eeq
where $\delta_{\beta}S$ is defined in (\ref{dS}). As a result we
obtain the equation
\beq \label{EZbd} \nonumber &&\frac{\partial
Z_{\beta}(J)}{\partial\beta}=\frac{i}{\hbar}
\Big[\frac{1}{2}\Big(\frac{\hbar}{i}\Big)^2\frac{\delta^2}{\delta
L^a\delta L^a}- \frac{\hbar}{2i}\frac{\delta}{\delta L^a}N^a
\big(\sfrac{\hbar}{i}\sfrac{\delta}{\delta K},
\sfrac{\hbar}{i}\sfrac{\delta}{\delta {\bar K}}\big)+\\
&&\qquad +\frac{1}{4}N^a\big(\sfrac{\hbar}{i} \sfrac{\delta}{\delta
K},\sfrac{\hbar}{i}\sfrac{\delta}{\delta {\bar K}}
\big)N^a\big(\sfrac{\hbar}{i}\sfrac{\delta}{\delta K},
\sfrac{\hbar}{i}\sfrac{\delta}{\delta {\bar K}}\big)+
m^2\Big(\frac{\hbar}{i}\Big)^2\frac{\delta^2}{\delta {\bar
K}^a\delta K^a}\Big] Z_{\beta}(J), \eeq
describing the dependence of Green's functions on the parameter $\beta$.
In terms of the generating functional of connected Green's functions,
$W_{\beta}(J)=\hbar/i \ln Z_{\beta}(J)$, the equation (\ref{EZbd})
takes the form
\beq \label{EWbd}
\nonumber
\frac{\partial W_{\beta}(J)}{\partial\beta}&=&
\frac{1}{2}\Big(\frac{\delta W_{\beta}}{\delta L^a}
\frac{\delta W_{\beta}}{\delta L^a}+
\frac{\hbar}{i}\frac{\delta^2 W_{\beta}}{\delta L^a\delta L^a}\Big)-\\
\nonumber &-&\frac{1}{2}\Big(\frac{\delta W_{\beta}}{\delta L^a}+
\frac{\hbar}{i}\frac{\delta }{\delta L^a}\Big) N^a\big(\sfrac{\delta
W_{\beta}}{\delta K}+ \sfrac{\hbar}{i}\sfrac{\delta}{\delta
K},\sfrac{\delta W_{\beta}}{\delta {\bar K}}+
\sfrac{\hbar}{i}\sfrac{\delta}{\delta {\bar K}}\big)+\\
\nonumber &+&\frac{1}{4}N^a\big(\sfrac{\delta W_{\beta}}{\delta K}+
\sfrac{\hbar}{i}\sfrac{\delta}{\delta K},\sfrac{\delta
W_{\beta}}{\delta {\bar K}}+ \sfrac{\hbar}{i}\sfrac{\delta}{\delta
{\bar K}}\big) N^a\big(\sfrac{\delta W_{\beta}}{\delta K}+
\sfrac{\hbar}{i}\sfrac{\delta}{\delta K},\sfrac{\delta
W_{\beta}}{\delta {\bar K}}+
\sfrac{\hbar}{i}\sfrac{\delta}{\delta {\bar K}}\big)+\\
&+&m^2\Big(\frac{\delta W_{\beta}}{\delta{\bar K}^a}
\frac{\delta W_{\beta}}{\delta K^a}+
\frac{\hbar}{i}\frac{\delta^2 W_{\beta}}{\delta{\bar K}^a\delta K^a}\Big).
\eeq
Introducing the generating functional of vertex functions (effective
action),  $\Gamma_{\beta}(\phi)$, being defined through the Legendre
transformation of $W_{\beta}(J)$,
\beq \Gamma_{\beta}(\phi)=W_{\beta}(J)-J_i\phi^i,\quad \phi^i=
\frac{\delta W_{\beta}}{\delta J_i},\quad
\frac{\delta\Gamma_{\beta}(\phi)}{\delta\phi^i}=-J_i, \eeq
the equation corresponding to (\ref{EWbd}) has the form
\beq \label{EGbd} \frac{\partial
\Gamma_{\beta}(\phi)}{\partial\beta}=
\frac{1}{2}{\hat B}^a{\hat B}^a-
\frac{1}{2}{\hat B}^aN^a\big({\hat C},{\hat{\bar C}}\big)+
\frac{1}{4}N^a\big({\hat C},{\hat{\bar C}}\big)
N^a\big({\hat C},{\hat{\bar C}}\big)+
m^2{\hat{\bar C}^a}{\hat C}^a,
\eeq
where the notations
\beq \label{EGbd1} {\hat\phi}^i=\phi^i+i\hbar (\Gamma^{``-1})^{ij}
\frac{\delta_l}{\delta\phi^j},\quad
(\Gamma^{''})_{ij}=\frac{\delta_l}{\delta\phi^i}
\Big(\frac{\delta\Gamma_{\beta}}{\delta\phi^j}\Big),\quad
(\Gamma^{``-1})^{ik}\Gamma_{kj}=\delta^i_j, \eeq
were used. We see that the dependence of effective  action on the
parameter $\beta$ does not disappear even on-shell defined by the
equations of motion of $\Gamma_{\beta}(\phi)$ that confirms a
physical character of the parameter $\beta$. In tree approximation
$\Gamma_{\beta}=S,\;{\hat\phi}^i=\phi^i$, and from (\ref{EGbd}) it
follows (\ref{dS}).

\section{Discussion}
We investigated the dependence of Green's  functions for the CF
model on the parameter resembling the gauge parameter in massless
Yang-Mills fields. In particular, it was shown that the effective
action for this model  depends on this parameter on-shell. It
allowed to consider this parameter as a physical one which can be
associated with definition of mass for scalar  anticommuting fields
of the CF model. It was found that violation of nilpotency of the
BRST symmetry can be interpreted as a source for appearance of an
additional physical parameter in comparison with a gauge theory for
which the full configuration space has the same structure. This
situation is quite similar to that in the Gribov-Zwanzider theory
\cite{Gribov,Zwanziger1,Zwanziger2} when violation of the BRST
symmetry of the Gribov-Zwanziger action was interpreted as a source
for the Gribov parameter to be a physical parameter \cite{VZ}.

It was pointed out that from the beginning  the CF model should be
considered as non-degenerated system of massive vector fields and
massive scalar anticommuting fields. From this point of view the
analysis of unitarity given in \cite{Kondo} looks like incomplete
because the physical state space should include particles
corresponding to massive scalar anticommuting fields as real ones.
In turn, presence of these particles in physical state space does
not give a chance for the CF model to be unitary because of the
breakdown of norm-positivity \cite{NOj,KO}.

\section*{Acknowledgments}
\noindent
The author thanks I.L. Buchbinder and I.V. Tyutin for
useful discussions of this paper.
The work is supported by  the LRSS
grant 224.2012.2  as well as by the RFBR grant 12-02-00121 and the
RFBR-Ukraine grant 11-02-90445.


\end{document}